\documentclass[pdflatex,sn-mathphys-num]{sn-jnl}

\setcitestyle{authoryear}

\usepackage{graphicx}%
\usepackage{multirow}%
\usepackage{amsmath,amssymb,amsfonts}%
\usepackage{fullpage}
\usepackage{caption}

\begin{document}

\title{Lorentzian correction for the evolution of the CMB temperature}

\author{\fnm{Artur} \sur{Novais}}\email{arturnovais@gmail.com}

\author{\fnm{A.L.B.} \sur{Ribeiro}}\email{albr@uesc.com.br}

\affil{\orgdiv{Laboratório de Astrofísica Teórica e Observacional}, \orgname{Universidade Estadual de Santa Cruz}, \orgaddress{\street{Rodovia Jorge Amado}, \city{Ilheus}, \postcode{45662-900}, \state{Bahia}, \country{Brazil}}}

\abstract{Observational evidence consistently shows that the universe is spatially flat and undergoes Lorentzian time dilation as a function of redshift. In combination, such discoveries suggest that a Minkowskian description of Cosmology might be technically viable. The thermal evolution that transpires in a conformal spacetime is herein derived. The description is constrained by the energy conservation of a unified cosmic fluid. The resulting model puts forth a Lorentzian correction for the temperature of the Cosmic Microwave Background Radiation as a function of redshift, which improves current data fitting without adding any free parameter. Furthermore, it sheds light upon the early galaxy formation problem: our model  predicts up to 0.86 Gyr older objects within the first two billion years of the structure evolution in the universe.}


\maketitle

\section{Introduction}\label{sec1}

The Cosmic Microwave Background (CMB) indicates that space possesses but a vanishing curvature on cosmological scales \citep{AghaninPlanck2018},\citep{Efstathiou2020TheEvidence}. Supernovae surveys \citep{RiessTimeDilation} and recent studies on primordial quasars \citep{Lewistimedilation} also bring forth evidence that Lorentzian time dilation is an observable effect at high recession speeds.
Such data suggest the possibility of a Minkowskian description of cosmology.

Over the course of the 20th century, cosmological models utilizing the Minkowskian background were developed as attempts at preserving the conformal quality of spacetime.
\cite{milne1933world} proposed a thought experiment where a distribution of particles endowed with arbitrary velocities \(0<v<c\) around any observer inevitably produces a radial expansion scenario governed by the Hubble law.  \cite{infeld1945new} generalized Milne´s results, deriving multiple cases of conformal universes embedded in the Minkowskian metric and their respective equations of motion. The authors demonstrated that such cases were geometrically equivalent to Friedmann-Lemaître-Robertson-Walker (FLRW) universes. 

Later, \cite{Tauber1967Expanding} explicitly solved Einstein's equations for the FLRW conformally flat-form metrics and for various types of equation of state.
\cite{Endean1994Redshift} considered transformations of the FLRW metrics in the case of open three-dimensional space curvature, and also for closed three-dimensional space curvature \citep{Endean1995Resolution}. Subsequently, \cite{Endean1997Cosmology} found a possible solution to cosmological age and redshift-distance difficulties by applying the appropriate conformally flat spacetime coordinates to the standard solution of the field equations in a standard dust model closed universe. 
\cite{Ibison2007OntheCOnformal} showed that the metrics of all RW models $(k = 0, \pm 1)$ are conformally flat;
and \cite{Romero2012Conformally} showed that any
RW cosmological model is totally determined by the Weyl scalar field $\phi$ while spacetime remains fixed --  $\phi$ may be considered as a gauging function determining the behaviour of clocks and measuring rods in a Minkowski spacetime. Finally, \cite{lombriser2023cosmology}
presents a formulation of cosmology in Minkowski spacetime, where the cosmological constant problem is absent. 

All these results reiterate the relevance of studying cosmology in Minkowski spacetime. The aim of the present work is to develop a Lorentz-invariant description of cosmology, which can be understood as a conformal transformation of the FLRW metric into the Minkowski space with a cosmic fluid undergoing Hubble flow from the perspective of any given inertial observer. This path leads us to a Lorentzian correction for the evolution of the CMB temperature.

\section{Hubble flow in Minkowski space}

The Hubble law defines the proportionality between distance and velocity:

\begin{equation}\label{(1)}
    v=H_0 r
\end{equation}

Considering that the resulting Hubble flow is subject to Lorentz transformations, one can determine the contracted length \(dx\):

\begin{equation}\label{(2)}
dx=dr\sqrt{1-\left( \frac{v}{c} \right)^2}
\end{equation}

Next, defining the Hubble radius \(R_H\) as the distance \(r\) where the recession velocity equals the speed of light, one obtains:

\begin{equation}\label{(3)}
c=H_0 R_H
\end{equation}

\[dx=dr\sqrt{1-\left( \frac{H_0 r}{H_0 R_H} \right) ^2 }\]

\begin{equation}\label{(4)}
dx=dr\sqrt{1-\left( \frac{r}{R_H} \right) ^2 }
\end{equation}

Thus, the transformed segment \(dx\) is expressed as a function of \(r\). For simplicity purposes, it is beneficial to define the angle \(\alpha\), such that:

\begin{equation}\label{(5)}
\sin{\alpha}\ \triangleq \frac{r}{R_H}
\end{equation}

And equation (4) becomes:

\[dx=dr\sqrt{1- \sin^2 \alpha\ }\]
\begin{equation}\label{(6)}
dx=dr\cos{\alpha}
\end{equation}

Finally, the sum of all consecutive segments \(dx\) yields the integrated distance \(x\), observed in the expanding Minkowskian substratum: 

\[x=\int_0^r dr\cos{\alpha}\]
\begin{equation}\label{(7)}
x=\frac{R_H}{2}(\sin{\alpha} \cos{\alpha}+\alpha)
\end{equation}

\begin{figure}
    \centering
    \includegraphics[width=0.7\linewidth]{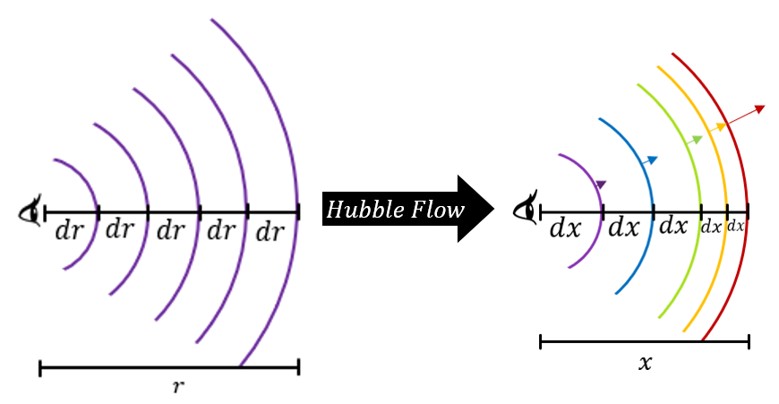}
    \caption{Length contraction of successive sections of the cosmic radius yielding a conformally transformed distance.}
\end{figure}

In this description, the length contraction does not affect the recession velocity, since time dilation is also present in Minkowski spacetime, i.e., as a signal crosses a contracted length, its local time passage \(dt_L\) is slowed down by the same factor and the original speed is maintained.

\begin{equation}\label{(8)}
v=\frac{dx}{dt_L}=\frac{dr\cos{\alpha}}{dt\cos{\alpha}}=\frac{dr}{dt}
\end{equation}

Therefore, the signal transmission is always conformal and instant velocities are preserved by the Lorentz transformations. This generates an important consequence for the Hubble flow equation, given that the speeds must be conserved when \(x\) replaces \(r\) as the observed distance. The result is a Lorentz-corrected Hubble parameter \((H_L)\).

The conformal condition equation is:

\begin{equation}\label{(9)}
H_0 r=H_L x
\end{equation}

Introducing the transformed radius \(x\) obtained in (7) and isolating \(H_L\):

\[H_L=H_0 \frac{2r}{R_H (\sin{\alpha} \cos{\alpha} +\alpha )}\]

Finally, using the definition of \(\sin{\alpha}\) given by (5):

\begin{equation}\label{(10)}
H_L=H_0 \frac{2\sin{\alpha}}{(\sin{\alpha }\cos{\alpha} +\alpha )}
\end{equation}

Such relation determines that, at any given time, the Lorentz-corrected Hubble parameter \(H_L\) is not constant for the entire cosmological radius, but gently ascends with the distance, i.e., with \(\alpha\) and the proximity to the horizon. \(H_0\), in turn, is a temporal function that continuously declines with the expansion of the universe, as \(R_H\) increases, as we can see by rearranging (3):

\begin{equation}\label{(11)}
H_0 (t)=\frac{c}{R_H(t)}
\end{equation}

Moreover, since the Hubble radius always expands at the speed of light:
\begin{equation}\label{(12)}
R_H=ct
\end{equation}

It is clear from this equation that the Minkowskian description of a unified cosmic fluid shall carry fundamental similarities to the \(R_h=ct\) model put forth by \cite{Melia2012Rh=ct}, albeit utilizing different metrics.

Substituting (12) in (11), one gets:

\begin{equation}\label{(13)}
H_0\ =\frac{1}{t}
\end{equation}

Therefore, in this work´s description, hereon named \textit{ZEUS} (Zero-Energy Unified Substratum), the age of the universe measured by the clock of the observer is always equal to the inverse of the Hubble constant \(H_0\) at that epoch. In consonance with the cosmological principle, this fundamental property is equally valid for all inertial observers at any given era, dismissing the need for a cosmic reference frame and resolving the present-time age coincidence problem \citep{Kutschera2007Coincidence}. Next, we present a new perspective on
the early structure formation problem
according to \textit{ZEUS}.

\begin{figure}
    \centering
    \includegraphics[width=0.6\linewidth]{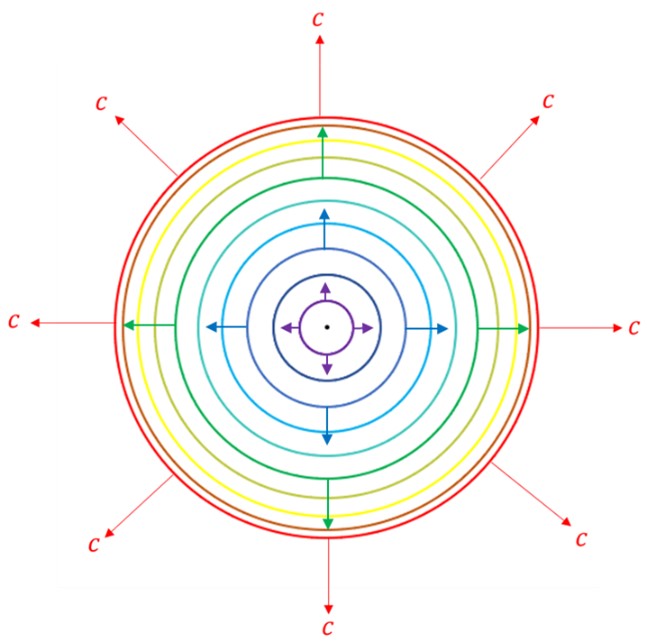}
    \caption{Expanding cosmic fluid in Minkowski space.}
    \captionsetup{justification=centering}
\end{figure}

\section{Time Dilation and a new perspective on the early galaxy problem}

From the definition of \(\sin{\alpha}\) and \(H_0\), it is straightforward to show that the Hubble flow does not alter the angle \(\alpha\) of a receding object over time, which means that the Lorentz contraction creates a fixed radial gradient in $\alpha-$space, which persists throughout cosmic expansion history.

\begin{equation}\label{(14)}
\frac{d\sin{\alpha}}{dt}=0
\end{equation}

In Minkowski coordinates, the total time dilation perceived by an observer receiving signals from a receding source is the combination of the Lorentzian \(\gamma\) – here treated as \(1/\cos{\alpha}\) – due to the velocity itself, and an extra factor \((1+\beta)\) – here expressed as \((1+\sin{\alpha})\) – due to the continuous increase in separation.

\begin{equation}\label{(15)}
\frac{dt_0}{dt_L}=\frac{(1+\sin{\alpha})}{\cos{\alpha}}
\end{equation}

One can also express the term \((1+z)\) as a function of \(\alpha\), arriving at the conclusion that the relativistic Doppler redshift is the exact manifestation of the time dilation.

\begin{equation}\label{(16)}
(1+z)= \frac{\sqrt{1+\frac{v}{c}}}{\sqrt{1-\frac{v}{c}}}=\frac{(1+\sin{\alpha})}{\cos{\alpha}}
\end{equation}

Observational evidence supports \((1+z)\) as the time dilation term. \cite{Davis2004Confusion} also concluded that a Lorentz-Minkowski type of expansion leads to this term, which is the same as the one given by the FLRW metrics. They proceeded, however, to calculate an incorrect luminosity distance, which was then rectified by \cite{Chodorowski2005Cosmology}, who showed that the magnitude-redshift diagram of a Minkowskian description is, in fact, remarkably close to the best \(\Lambda\)CDM fit.

A conundrum of crescent notoriety facing current Cosmology is the observation of complex structure - quasars and mature galaxies - at precocious epochs of the cosmic evolution. This rapid emergence challenges the known structure formation mechanisms given by hierarchical models and appears to contradict theoretical constraints, such as the Eddington limit for black hole accretion (\citep{Robertson2023IntenseStarFormingGalaxies}, \citep{Melia2014PrematureFormation} and \citep{Melia2015Supermassive}).

The root of the conflict is the correlation between the redshift and the cosmic age given by the \(\Lambda\)CDM framework, i.e., if another valid cosmological description adjusts the age of the structures at the instant of emission, providing more time for them to have formed at the observed redshifts, the problem may be solved.

\begin{figure}
    \centering
    \includegraphics[width=0.7\linewidth]{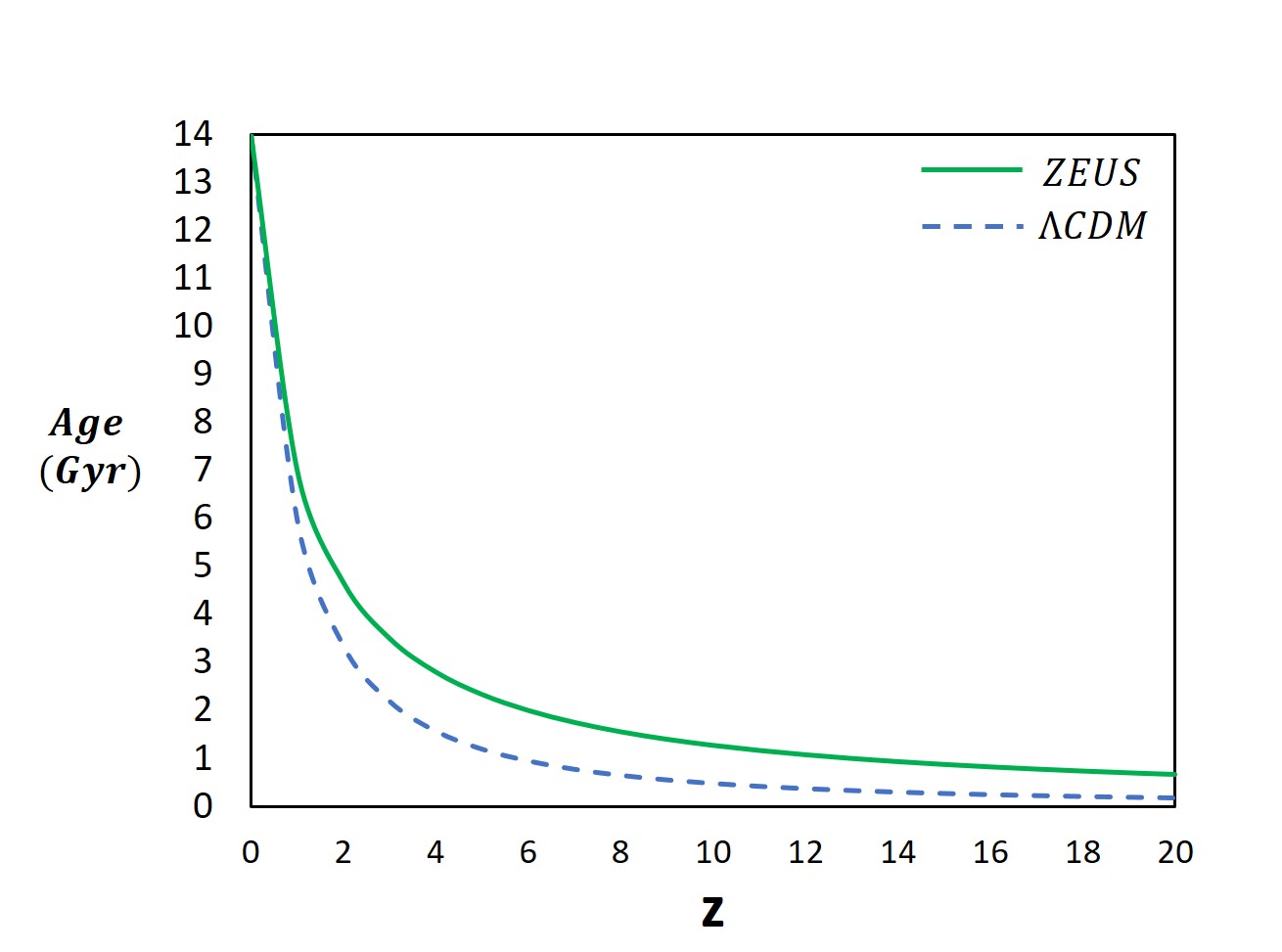}
    \caption{Time available for the evolution of a given source versus its redshift considering the time dilation effect. Solid curve is the ZEUS prediction and dashed curve is the $ \Lambda$CDM correlation.}
\end{figure}

\begin{figure}
    \centering
    \includegraphics[width=0.7\linewidth]{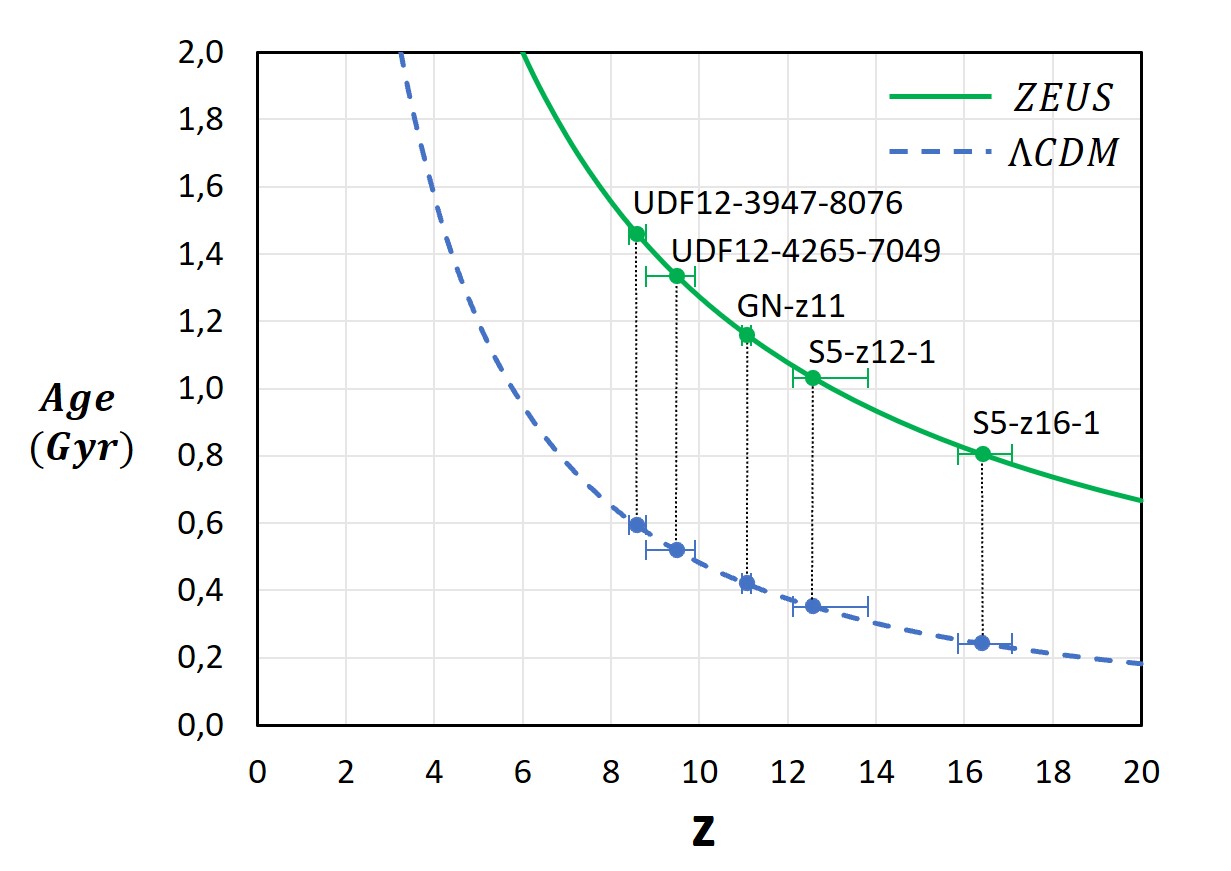}
    \caption{Same as figure 3 but focusing upon the first two billion years of structure evolution. Observed galaxies  from HST (\citep{Oesch2016GNZ11}, \citep{Ellis2013TheAbundance}) and JWST (\citep{Harikane2023}), are projected onto $ \Lambda$CDM and ZEUS timelines.}
\end{figure}

In the \textit{ZEUS} description, since the redshift of a given source is fixed, it constitutes a constant in any time integration, hence this very factor is applicable to vast cosmic eras as well as infinitesimal intervals. Therefore, the age of the object in the \textit{ZEUS} model is calculated by dividing the current age of the universe, here considered 14 Gyr, by the factor $(1+z)$.

\begin{equation} 
\Delta t_{source}=\frac{\Delta t_{observer}}{(1+z)} \nonumber
\end{equation}

\begin{equation}\label{(17)}
Age_{source}=\frac{Age_{localUniverse}}{(1+z)}
\end{equation}

In contrast, the \(\Lambda\)CDM evolution presents transitory phases that render the age-redshift correlation more complex and time-compressed, as shown in Figure 3. Figure 4 then zooms into the first two billion years of the structure evolution in order to highlight the significant difference in the age-versus-redshift curves. We show objects observed by HST and JWST that are 0.56 to 0.86 Gyr older in \textit{ZEUS} than in the \(\Lambda\)CDM model. The early galaxy formation problem is alleviated within the \textit{ZEUS} description for it grants more time for the galaxies and quasars to develop.

\section{Energy Conservation}

In this section, the dynamics of the \textit{ZEUS} model is explored. It is based upon the hypothesis that the total energy of the unified cosmic substratum is always zero. 
It is important to emphasize that the universe is not
filled with a single fluid, but the mixture of radiation,
baryons, and some manifestation of dark matter and dark energy, evolving under the zero energy condition:

\begin{equation}\label{(18)}
\sum \rho_i + 3\sum p_i =0
\end{equation}
			
This premise is similar to the unified medium with zero active mass and cosmic equation of state $ p/\rho = -1/3$ proposed by \cite{Melia2012Rh=ct} in the FLRW metric. Here, however, the Lorentzian correction applied to the Hubble flow in Minkowski spacetime leads to an energy density gradient along cosmic distances which is more descriptive of some CMB features, e.g., the high entropy density and the temperature evolution with redshift (see Section V).

First, from a mechanical perspective, the relativistic kinetic energy, written in terms of \(\alpha\), is given by:

\[K=m_0c^2\left( (\frac{1}{\sqrt{1-(v/c)^2}}-1\right)=\frac{m_0v^2}{sin^2\alpha} \left( \frac{1}{cos\alpha}-1\right)\]

\begin{equation}\label{(19)}
K=\frac{m_0v^2}{\cos{\alpha}(1+\cos{\alpha})}
\end{equation}

Next, introducing the relativistic potential energy produced by the gravitational field of a sphere centered at the inertial observer with transformed radius \(x\) and average energy density \(\rho\):

\begin{equation}\label{(20)}
P=-\frac{8\pi G\rho x^2m_0}{3c^2\cos{\alpha}(1+\cos{\alpha})}
\end{equation}

In agreement with the observed spatial flatness, the total mechanical energy of free particles that move exclusively due to the Hubble flow is considered to be zero:

\[\frac{m_0v^2}{\cos{\alpha}(1+\cos{\alpha})}-\frac{8\pi G\rho x^2m_0}{3c^2\cos{\alpha}(1+\cos{\alpha})}=0\]

\begin{equation}\label{(21)}
v^2=\frac{8\pi G\rho x^2}{3c^2}
\end{equation}

Next, expressing the Hubble law:

\[H_L^2x^2=\frac{8\pi G\rho x^2}{3c^2}\]
\begin{equation}\label{(22)}
H_L^2=\frac{8\pi G\rho }{3c^2}
\end{equation}

This result is analogous to the first Friedmann equation for a flat space with total energy density equal to the critical value and vanishing cosmological constant. The key distinction is that the Minkowskian coordinates produce a Lorentz-corrected Hubble parameter \( H_L\), which slightly grows with distance. This means that the average energy density \(\rho\) of the cosmic sphere also presents a radial gradient at any given point in time. This can be demonstrated by taking (22) and substituting the \(H_L\) obtained in (10):

\[H_0^2 \frac{4\sin^2\alpha}{(\sin{\alpha}\cos{\alpha }+\alpha)^2}=\frac{8\pi G\rho }{3c^2}\]

\begin{equation}\label{(23)}
\rho=\frac{3c^2}{8\pi G}H_0^2\frac{4\sin^2\alpha}{(\sin{\alpha} \cos{\alpha} +\alpha)^2}
\end{equation}

At the limit \(\alpha \rightarrow 0\), where the small-angle approximation (\(\sin{\alpha}\rightarrow\alpha\) and \(\cos{\alpha}\rightarrow1\)) is applicable, one can calculate the energy density in the spatial vicinity of the observer:

\begin{equation}\label{(24)}
\rho_0=\frac{3c^2}{8\pi G}H_0^2
\end{equation}

\begin{equation}\label{(25)}
\rho=\rho_0 \frac{4\sin^2\alpha}{(\sin{\alpha} \cos{\alpha }+\alpha)^2}
\end{equation}

Note that this expression determines the average density of the entire cosmic sphere from \(\alpha=0\) up to an \(\alpha\) of interest. The differential energy density \(\varepsilon\) at \(\alpha\) itself is the increment of the total energy of the sphere \(dU\) with an increment of the volume \(dV\).

\begin{equation}\label{(26)}
V=\frac{4}{3} \pi x^3
\end{equation}

Applying the expression of \(x\) obtained in (7):

\begin{equation}\label{(27)}
V=\frac{\pi}{6} R_H^3(\sin{\alpha} \cos{\alpha} +\alpha)^3
\end{equation}

And differentiating with respect to \(\alpha\):

\begin{equation}\label{(28)}
\frac{dV}{d\alpha}=\pi R_H^3\cos^2\alpha(\sin{\alpha} \cos{\alpha} +\alpha)^2
\end{equation}

Next, defining the internal energy \(U\):

\begin{equation}\label{(29)}
U=\rho V
\end{equation}

Which can be calculated by employing Equations (23) and (27) for \(\rho\) and \(V\):

\begin{equation}\label{(30)}
U=\frac{2}{3}\pi R_H^3\rho_0 \sin^2\alpha(\sin{\alpha} \cos{\alpha} +\alpha)
\end{equation}

And differentiating with respect to \(\alpha\):

\begin{equation}\label{(31)}
\frac{dU}{d\alpha}=\frac{4}{3}\pi R_H^3\rho_0 \sin{\alpha }\cos{\alpha}(2\sin{\alpha} \cos{\alpha} +\alpha)
\end{equation}

One may finally determine the differential energy density \(\varepsilon\) at an \(\alpha\) of interest:

\begin{equation}\label{(32)}
\varepsilon=\frac{dU}{dV}=\frac{dU/d\alpha}{dV/d\alpha}
\end{equation}

\begin{equation}\label{(33)}
\varepsilon=\rho_0 \left[ \frac{4}{3}\frac{\sin{\alpha}(2\sin{\alpha} \cos{\alpha} +\alpha)}{\cos{\alpha}(\sin{\alpha} \cos{\alpha} +\alpha)^2}\right]
\end{equation}

For notation simplicity, the term in square brackets is hereon denoted by \(\xi\):

\begin{equation}\label{(34)}
\xi=\frac{4}{3}\frac{\sin{\alpha}(2\sin{\alpha} \cos{\alpha }+\alpha)}{\cos{\alpha}(\sin{\alpha} \cos{\alpha} +\alpha)^2}
\end{equation}

\begin{equation}\label{(35)}
\varepsilon=\rho_0 \xi
\end{equation}

Such gradient may have measurable implications for the temperature of the Cosmic Microwave Background, as studied in the next section. 

\section{Temperature of the CMB}

The linear \(T_{CMB}(z)\) is a property of standard adiabatic models that presuppose homogeneity and isotropy, such as \(\Lambda\)CDM.

\begin{equation}\label{(36)}
    T_{CMB}(z)=T_0(1+z)
\end{equation}

Such proportionality is not bound to a particular metric theory when assuming  that the cosmos expands isotropically, photon has no mass, the CMB radiation is thermal and the first law of thermodynamics is true \citep[e.g.][]{abitbol2020measuring}.
However, a departure from linearity 
would require important and hard to detect
distortions in the Planck spectrum of the CMB \citep{chluba2014tests}, which could be used to constrain
alternative scenarios, e.g., cases where photons are either created or destroyed, as explored by \cite{lima2000radiation}, or 
modifications of gravity via the
presence of a scalar field with a multiplicative coupling to the electromagnetic Lagrangian \citep{hees2014breaking}.  
 At present, there appears to be no inconsistency of the $T_{CMB}(z)$ data with the $\Lambda$CDM model \citep[e.g.][]{arjona2020machine}.
An extensive program of experiments based on new technologies will be able to detect
minimal  distortions in the energy spectrum of the CMB in the near future
\citep{kogut2019cmb}. 

A different approach to study the evolution of the CMB temperature is to introduce a Lorentzian correction in the $T_{CMB}(z)$ function. In this case, the universe would be strictly flat, with a unified fluid describing its contents at all times. The aim of this work is to show that a Minkowskian description of cosmology,
where the thermal evolution takes place
in a conformal spacetime, can improve the
data fitting of $T_{CMB}(z)$ for current datasets. This has important consequences for flatness tests and the foundations of cosmology.

While each component of the \(\Lambda\)CDM model behaves in a particular manner in terms of \(\rho_i(t)\), in the \textit{ZEUS} description, the energy density of the unified cosmic fluid possesses a universal behavior: it decreases with the square of the proper time (\(\rho_0\propto t^{-2}\)). The second difference is the spatial gradient, which we characterized in the previous section with the term \(\xi\).

\[\varepsilon=\rho_0(t)\xi(\alpha)\]
\begin{equation}\label{(37)}
\varepsilon=\frac{3c^2}{8\pi G}\frac{1}{t^2}\xi(\alpha)
\end{equation}

Equipped with this cosmic energy profile across time and space, one can now tell the story of the CMB from the \textit{ZEUS} model´s perspective, including its interaction with structures at their respective redshifts and the resulting perception of the observer, once time dilation is taken into account.

In the beginning (\(t\sim 0\)), the temperature of the unified cosmic fluid is far too great around the observer (and even greater at higher values of \(\alpha\)) for any nucleon to form. One can directly express the temperature as a function of time by invoking the Stefan-Boltzmann law.

\[T=(\sigma \varepsilon)^{1/4}\]

\[T=(\sigma \rho_0)^{1/4} \xi^{1/4}\]

\begin{equation}\label{(38)}
T=T_0 \xi^{1/4}
\end{equation}

where \(\xi\) is a spatial function and \(T_0\) is a temporal function:

\begin{equation}\label{(39)}
T_0(t)=(\sigma \rho_0)^{1/4}=\left( \sigma \frac{3c^2}{8 \pi G}\right)^{1/4} \frac{1}{t^{1/2}}
\end{equation}

Therefore, the local temperature decreases monotonically. With time, the local temperature of the cosmic fluid eventually drops sufficiently to enable the primordial nucleosynthesis. Given yet more time, the plasma decoupling also takes place. However, since the universe was far smaller and the photons always travel conformally in Minkowskian coordinates, those first local photons are not the ones received at the present time.

Referring to equation (38), for every value of \(T_0\) there can be found a value of \(\xi^{1/4}\) that produces the temperature at which recombination takes place (any arbitrary value is possible, given that \(\xi\) tends to infinity at the horizon). As \(T_0\) decreases monotonically over time, this Surface of Last Scattering (SLS) must occupy greater and greater values of \(\xi\), i.e., it advances ever closer to the horizon.

Next, to understand how the CMB radiation interacts with the intervening galaxies between the SLS and the observer, as well as how the interaction signal is measured, two factors must be considered. First, as discussed in the previous section, the observed energy density of a given region of the cosmic fluid is affected by the total Lorentzian time dilation, which inserts a \((1+z)^{-1}\) factor to each time contribution. Second, in the CMB analysis, the primary emission surface (the SLS) is far out close to the horizon, meaning that any galaxy travels towards the CMB photons as it recedes from the observer, i.e., a blueshift is expected in this interaction when compared to the CMB energy density perceived by the static observer. This blueshift can be readily determined, since it is the inverse of the observed galaxy´s redshift. Such mechanism adds yet another \((1+z)^{-1}\) factor to the proper time. Hence, the total \(\rho_0\) time correction now becomes:

\begin{equation}\label{(40)}
\rho_{0CMB}\propto\frac{3c^2}{8 \pi G} \left[ \frac{(1+z)(1+z)}{t} \right]^2
\end{equation}

\[\rho_{0CMB}\propto\rho_0(1+z)^4\]

\begin{equation}\label{(41)}
\varepsilon_{CMB}\propto\rho_0(1+z)^4\xi
\end{equation}

\[T_{CMB}=(\sigma \varepsilon_{CMB})^{1/4}\]
\[T_{CMB}=(\sigma \rho_0)^{1/4}(1+z)\xi^{1/4}\]

\begin{equation}\label{(42)}
T_{CMB}=T_0(1+z)\xi^{1/4}
\end{equation}

\begin{figure}
    \centering
    \includegraphics[width=0.7\linewidth]{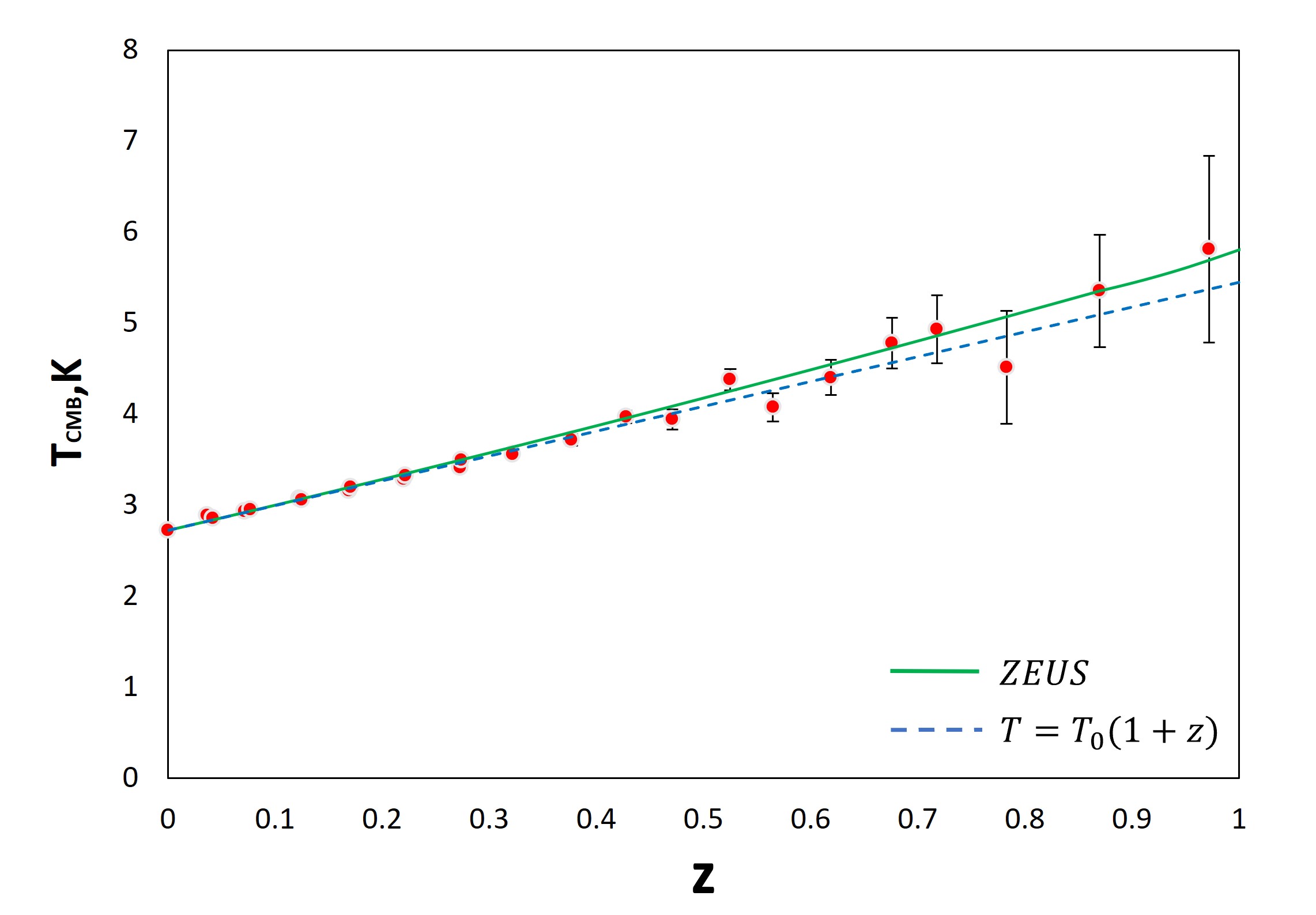}
    \caption{Circles are the results from \cite{Hurrier2014MeasurmentofTCMB} and \cite{deMartino2015Constraining} based on Planck map data and SZ effect at $z<1$. Dashed line shows the FLRW linear prediction with $\chi^2_{FLRW}=0.121$. Solid curve is the  ZEUS prediction with $\chi^2_{ZEUS}=0.118$.}
\end{figure}

\begin{figure}
    \centering
    \includegraphics[width=0.7\linewidth]{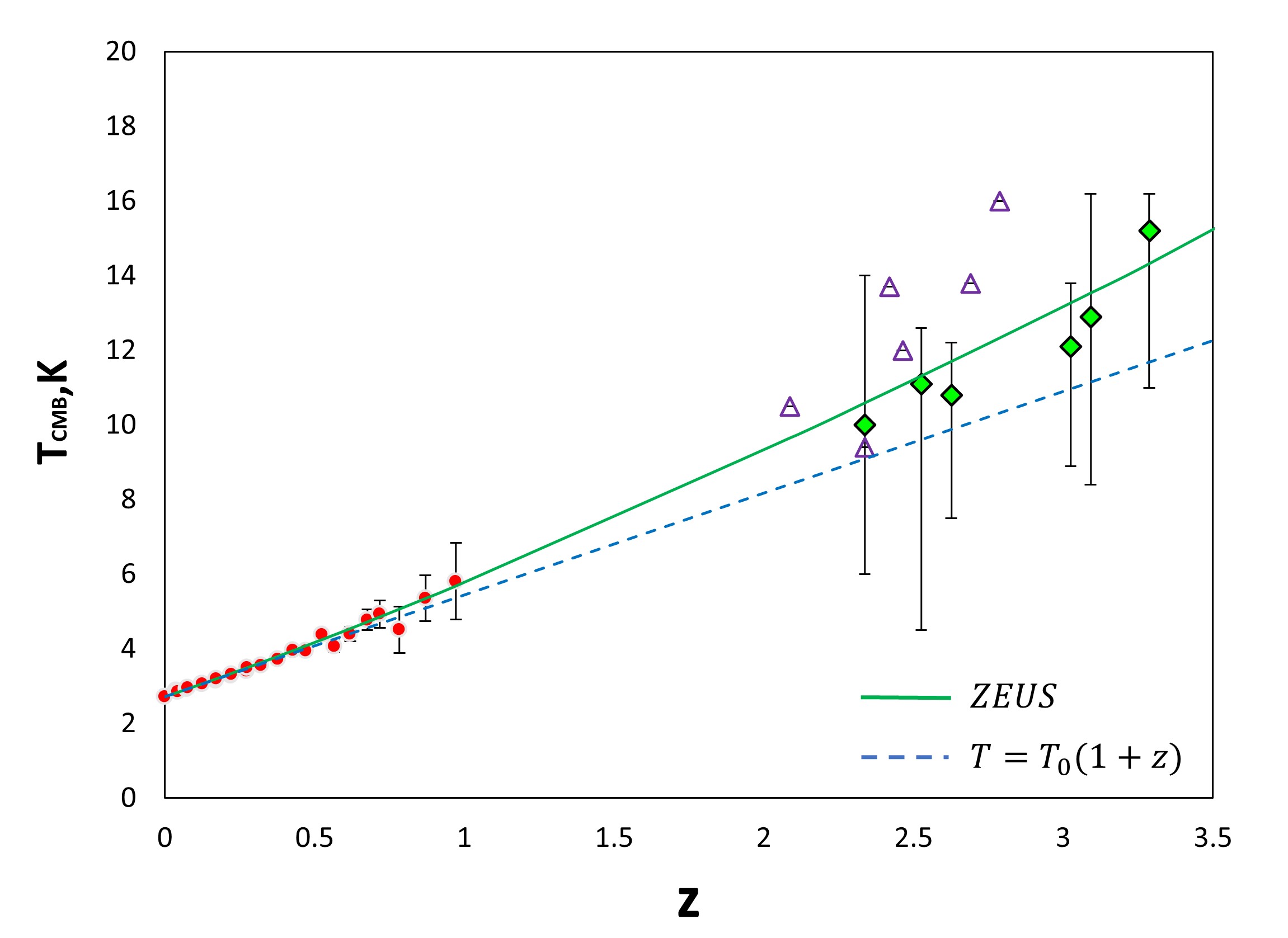}
    \caption{Same as figure 5 plus literature data based on Carbon fine-structure absorption lines represented as diamonds at $2<z<3.5$ (\citep{Srianand2000},\citep{Molaro2002},\citep{Balashev2010},\citep{Jorgenson2010},\citep{Guimaraes2012}). Triangles are upper bounds, not taken into account for $\chi^2$. $\chi^2_{FLRW}=1.636$ and $\chi^2_{ZEUS}=0.431$.}
\end{figure}

\begin{figure}
    \centering
    \includegraphics[width=0.7\linewidth]{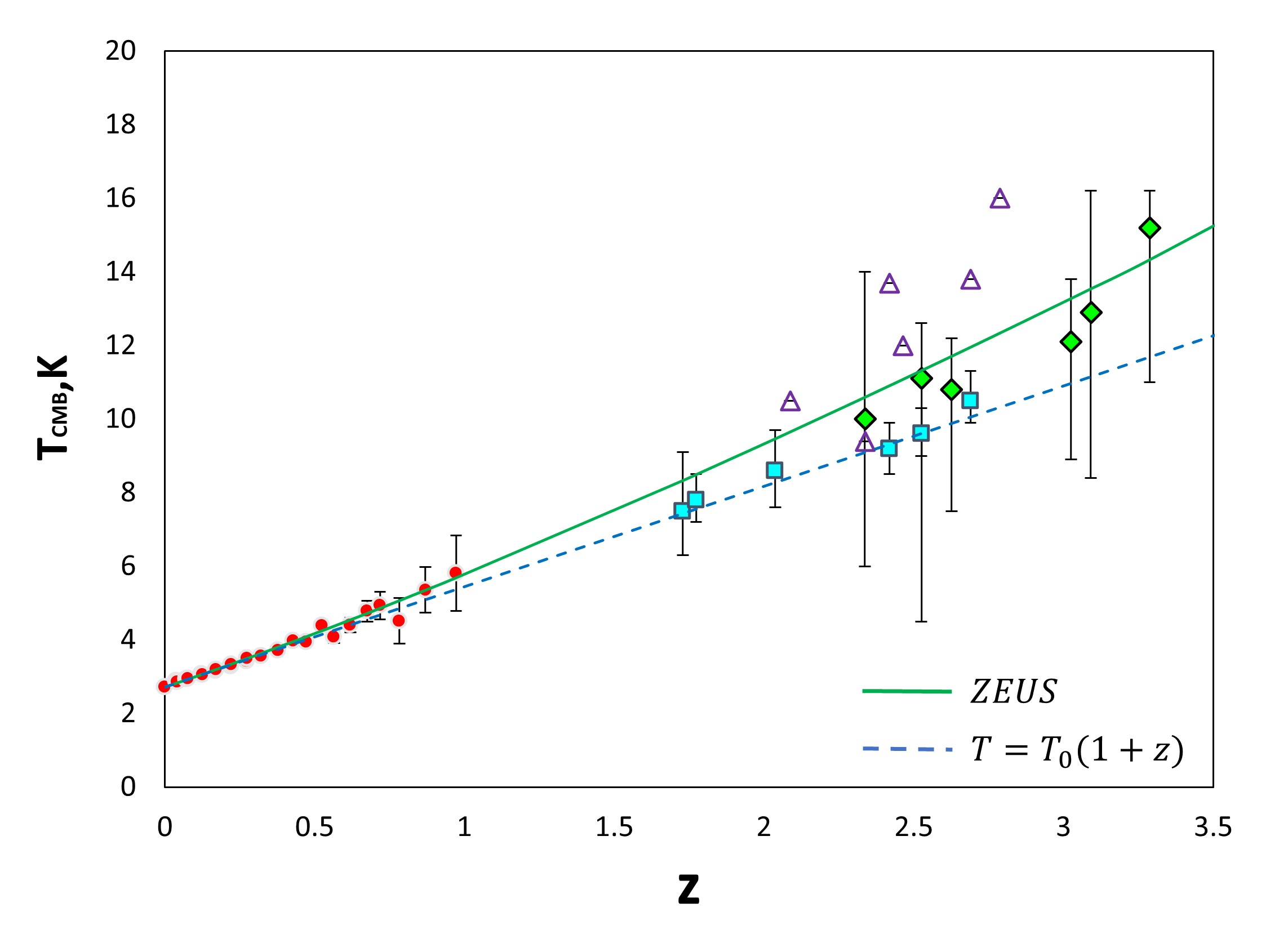}
    \caption{Same as figure 6 plus literature data based on CO rotation excitation represented as squares (\citep{Srianand2008First},\citep{Noterdaeme2010},\citep{Noterdaeme2011}). $\chi^2_{FLRW}=1.676$. $\chi^2_{ZEUS}=1.500$.}
\end{figure}

\begin{figure}
    \centering
    \includegraphics[width=0.7\linewidth]{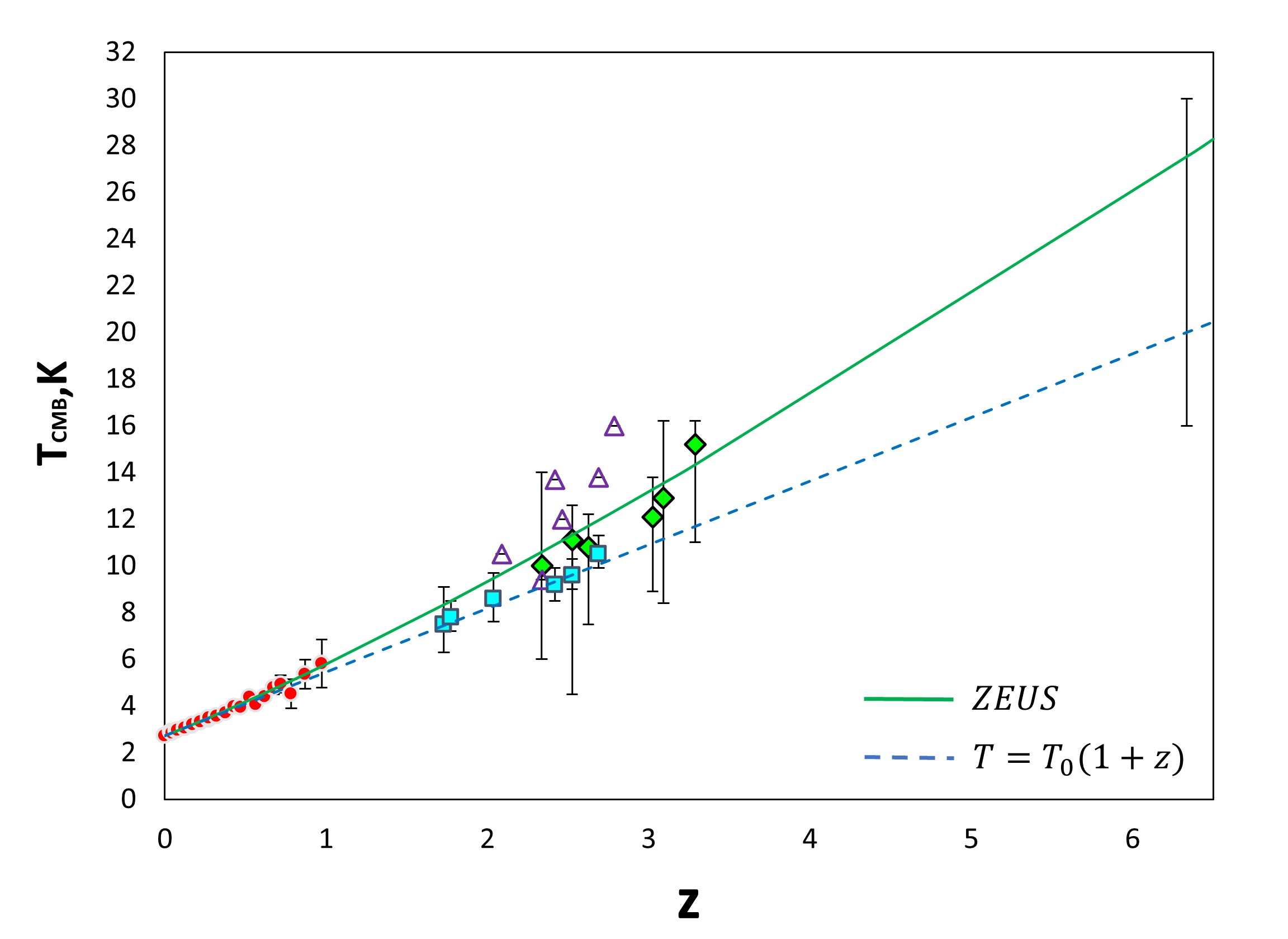}
    \caption{$T_{CMB}(z)$ $(0<z<6.5)$, including the result obtained by \cite{Riechers2022} via $H_2O$ absorption line.}
\end{figure}

It is remarkable that alternative perspectives of Cosmology, based on different coordinates, timelines, redshift interpretations and photonic histories, provide such similar predictions for the \(T_{CMB}(z)\). This reinforces the idea that, in a flat universe, which is clearly endowed with time dilation, the Minkowskian description of cosmological phenomena should be technically viable.

From the observational standpoint, at low redshifts \((0<z<1)\), the thermal Sunyaev-Zeldovich (tSZ) effect can be employed as a cosmic thermometer (see \citep{SunyaevZeldovich1970Small}, \citep{Fabbri1978Measurement},\citep{Raphaeli1980OnTheDetermination}). Figure 5 shows that the FLRW linear prediction and the \textit{ZEUS} curve are extremely similar along this range and both are good fits to the presently available datapoints.

As more profound redshifts are probed, different estimation techniques are required. Multiple studies rely upon atomic and molecular fine-structure levels observed in the absorption spectra of quasars. This method depends upon free parameters associated with local physical conditions, such as kinetic temperature, UV background intensity, gas number density, collisional corrections, etc \citep[e.g.][]{Klimenko2020Estimation}.

In order to assess the predictive potential of the \textit{ZEUS} description, we first gathered in figure 6 the data sources in the redshift range \(2<z<3.5\) that utilize atomic Carbon fine-structure levels. Despite the large error bars, early signs of a tendency may be observed: at higher redshifts, the datapoints tend to land above the \(T_0(1+z)\) line. As a result, the \textit{ZEUS} fit (\(\chi^2=0.431\)) is even better than the FLRW linear relation  (\(\chi^2=1.636\)).

Although the case of CO rotational levels excitation poses a greater statistical challenge, due to the low probability of detection and the ongoing debate on the required corrections, we incorporated the six results originally obtained by \cite{Srianand2008First}, \cite{Noterdaeme2010} and \cite{Noterdaeme2011} in Figure 7. The cumulative result shows that the \textit{ZEUS} fit (\(\chi^2=1.500\)) has good potential when compared with the FLRW fit (\(\chi^2=1.676\)).

It is worth noting that other studies further correct the inferred \(T_{CMB}(z)\) for the six CO datapoints, reasserting the sensitivity with respect to the assumptions about local physical conditions. For instance, \cite{Klimenko2020Estimation} present adjusted temperatures, deriving a fit that is closer to the FLRW linear prediction. \cite{Maeder2017} further reduces the \(T_{CMB}\) estimates by assuming additional galactic corrections, rendering both FLRW and \textit{ZEUS} utterly unfit. Clearly, a better grasp on the CO rotational levels methodology must be achieved. This includes a more profound understanding of local physical conditions as well as the detection of statistically robust samples.

Finally, \cite{Riechers2022} reported a 1\(\sigma\) range measurement of the CMB temperature at \(z=6.34\). Using observations of submillimetre line absorption from the \(H_2O\) molecule, they arrived at a temperature estimation of \(16.4-30.2K\). Most of the range lies above the above the \(T_0(1+z)\) line and, as seen in Figure 8, both \textit{ZEUS} and FLRW curves are able to accommodate the implied temperatures.

Larger datasets and greater refinement are required to expand the study. This work aims to showcase the fitting potential of a valid cosmological description in a developing observational field. The objective is to enrich future discussions, especially when it comes to the model-dependent aspects of the data treatment.

\section{Conclusion}

For over a century, the Minkowskian coordinates have been associated with specific motions of particles in restricted local scales. And correctly so: in a universe filled with energy and a spectrum of density fluctuations, one has no right to postulate \textit{a priori} that a special case of flat space is applicable to cosmological scales. However, the scenario shifts in light of observational evidence, which reveals a vanishing global curvature.

In a coherent development, time dilation at high recession speeds (high z) has been empirically verified. If correctly studied, this temporal transformation is shown to exactly match the Lorentzian type for moving sources.

Put together, such discoveries indicate that a Minkowskian description of cosmological phenomena should be technically viable. The question that remains is: can this description account for the evolution of the universe or is it limited to a kinematic snapshot of instant motions?

In the \textit{ZEUS} description, the unified dynamics is granted by the postulate of total energy conservation. From this assumption (fully integrated with flatness), the thermal evolution of the universe is derived, and every instant of proper time can be described in a Lorentz-invariant framework.

In this introduction to a novel cosmological description, for pedagogical and conciseness purposes, the focus was set upon two macro-observables: firstly, the widely discussed problem of the early galaxies in \(\Lambda\)CDM emerging from JWST and HST data, which is mitigated by \textit{ZEUS}. And secondly, the developing observational field of $T_{CMB}(z)$, still less discussed in the literature, and for which this work presents a unique, falsifiable and unprecedented approach with satisfactory fit to the current data, thus enriching the field´s dialogue as it grows.

Nonetheless, before declaring the model as viable, a series of tests must be conducted where the current standard cosmology is already successful, e.g., Supernovae and Gamma-ray Bursts studies, Baryonic Acoustic Oscillations tests, structure formation modelling and Big Bang Nucleosynthesis, as well as presenting solutions to current conflicts, such as the Hubble tension, the initial entropy problem and dark sector candidates.

\bmhead{Acknowledgements}
The authors are grateful to the referee for a detailed and constructive review, which has led to significant improvements in the manuscript. The authors also thank A. Kandus and G. Monerat for the useful suggestions. ALBR thanks CNPq, grant 316317/2021-7 and FAPESB INFRA PIE 0013/2016 for the support.

\nocite{*}

\bibliography{sn-bibliography}

\end{document}